\documentclass[twocolumn,aps,prl,superscriptaddress,tightenlines]{revtex4-1}

\usepackage{amsmath, amsfonts, amsthm, amssymb, graphicx, color}
\usepackage{hyperref}
\usepackage{subfigure}

\allowdisplaybreaks[3]
\def \bal#1\eal  {\begin{align} #1 \end{align}}
\newcommand{\be} {\begin{equation}}
\newcommand{\ee} {\end{equation}}
\newcommand{\nn} {\nonumber}

\newcommand{\pd} {\partial}

\def\ba{\begin{eqnarray}}
\def\ea{\end{eqnarray}}

\def\M{\mathcal{M}}

\def\({\left(}
\def\){\right)}

\def\p{\partial}

\begin{document}

\title{Positivity Bounds for Scalar Theories}
\author{Claudia de Rham}
\email{c.de-rham@imperial.ac.uk}
\author{Scott Melville}
\email{s.melville16@imperial.ac.uk}
\author{Andrew J.~Tolley}
\email{a.tolley@imperial.ac.uk}
\author{Shuang-Yong Zhou}
\email{Shuangyong.Zhou@imperial.ac.uk}
\affiliation{Theoretical Physics, Blackett Laboratory, Imperial College, London, SW7 2AZ, U.K. }
\date{\today}

\begin{abstract}

Assuming the existence of a local, analytic, unitary UV completion in a Poincar\'{e} invariant scalar field theory with a mass gap,  we derive an infinite number of positivity requirements using the known properties of the amplitude at and away from the forward scattering limit.
These take the form of bounds on combinations of the pole subtracted scattering amplitude and its derivatives. In turn,
these positivity requirements act as constraints on the operator coefficients in the low energy effective theory. For certain theories these constraints can be used to place an upper bound on the mass of the next lightest state that must lie beyond the low energy effective theory if such a UV completion is to ever exist.
\end{abstract}

\maketitle
The physical requirements of unitarity, locality, and crossing symmetry, are well known to provide powerful constraints on the scattering matrix of a Lorentz invariant theory, and were an integral part of the S-matrix program \cite{Martin:1969book,Eden:2012}. Relativistic locality and causality is encoded in the twin requirements of analyticity of the scattering amplitude, and polynomial boundedness. Taken together, these allow us to express the scattering amplitude in terms of dispersion relations with a finite number of subtractions, from which it is possible to infer bounds on the growth of the scattering amplitude at high energies.

It is only more recently that these constraints have been used to infer properties of low energy effective field theories (LEEFT) \cite{Adams:2006sv}. In doing so we assume the existence of a (possibly unknown) local Lorentz invariant UV completion and use its properties  to infer properties of the LEEFT. These typically come in the form of `positivity bounds', i.e. bounds on the sign of coefficients in the Wilsonian effective action. For example, it is known that for analytic $2$-to-$2$ scattering amplitudes in the forward scattering limit, an expansion in powers of the invariant mass $s$ must have positive coefficients \cite{Adams:2006sv}. It has also been suggested that these can be pushed away from the forward limit \cite{Pennington:1994kc,Vecchi:2007na,Manohar:2008tc,Nicolis:2009qm,Bellazzini:2014waa}.

Exploiting unitarity, analyticity and crossing symmetry of the full (unknown) UV complete theory, we will use the known properties of the scattering amplitude of a scalar theory at and away from the forward limit to show that there are an infinite number of such bounds on the pole subtracted scattering amplitude $B(s,t)$. These translate into bounds on the coefficients of every non-redundant (not removable by a field redefintion) operator that contributes to the $2$-to-$2$ scattering amplitude at tree level. We first derive the bounds on the exact quantum scattering amplitude, and then show how they may be applied to the tree-level amplitudes in the LEEFT. In certain cases, we will show how these constraints lead to  an upper bound of the mass of the first state that necessarily lies beyond the regime of validity of the LEEFT. \\

%%%%
\noindent{\bf Unitarity:}
The $2$-to-$2$ scattering amplitude is best expressed in terms of the Mandelstam variables \cite{Mandelstam:1958xc}:
$s$, the center of mass energy, $t$, the momentum transfer, related to the scattering angle by $\cos \theta = 1 + \frac{2t}{s - 4m^2}$, and their conjugate variable $u = 4m^2 - s -t$.
In order to derive positivity bounds on the scattering amplitude $A(s,t)$, we make use of unitarity in the form of the optical theorem,
${\rm Im}[A(s,0)] = \sqrt{s(s-4m^2)} \sigma(s) $, together with its implications for the partial wave expansion,
\be
A(s,t) = 16\pi \sqrt{\frac{s}{s-4 m^2}} \sum_{l=0}^{\infty} (2 l+1) P_l(\cos \theta)  a_l (s) \,,
\ee
namely $\text{Im} \, a_l (s) =  |  a_l (s) |^2 + \ldots    ~$,
where the omitted terms are proportional to cross section contributions from inelastic processes, which are positive. Unitarity therefore tells us that,
\begin{equation}
0 \leq | a_l (s) |^2  \leq  \text{Im} \, a_l (s)  \leq 1    \,, \quad {\rm for}\quad s \ge 4m^2\,.
\label{eqn:unitarity_al}
\end{equation}
Since the Legendre polynomials satisfy $\p^n_t P_l (1+t)  |_{t=0}  \geq 0 $,
it follows from \eqref{eqn:unitarity_al} that,
\be
\frac{\pd^n}{\pd t^n}{\rm Im} [A(s,t)] \Big|_{t=0} >0   \quad \forall \,\, n \geq 0  \quad {\rm and}\quad s \ge 4m^2 \,  .
\label{eqn:dImA}
\ee
This is a strict positivity for all $n$, since equality is only reached in the case of a free theory\;\footnote{Strictly speaking, the argument provided here only establishes $\partial_t^n {\rm Im} [A(s,0)] \ge 0 $. Strict positivity can nonetheless be rigorously proven, the details of which can be found in \cite{dRMTZ2:2017}.}  \\

%%%%
\noindent {\bf Analyticity}:
It is usually postulated that the amplitude $A(s,t)$ is analytic in the whole complex Mandelstam plane, except for poles and branch cuts implied by unitarity \cite{Mandelstam:1958xc} and crossing symmetry. However, many of the consequences of Mandelstam's proposal can be obtained by assuming much weaker provable analyticity conditions \cite{Martin:1962rt, Jin:1964zza}, which imply that $A(s,t)$ is analytic in the disk $|t| < 4m^2$ for fixed $s$ and in the twice cut plane of $s$ for fixed $t$ (excluding, of course, the possible poles of $s$ and $t$) \cite{Bros:1964iho,Martin:1965jj}. This is consistent with the statement that the scattering amplitude is analytic (modulo poles) in the Mandelstam triangle, $0 \le s,t,u < 4m^2$.

In a scalar theory, we expect the amplitude to have a simple $t$-channel pole at $t=m^2$, whose residue is $s$ independent (since it is determined by the on-shell vertex function which for a scalar theory can contain no remaining momentum dependence). This residue is necessarily real, and so $\text{Im} A (s,t)$ has no poles at  $t=m^2$. This implies that $\text{Im} A (s,t)$ is analytic with no poles in the region $|t|<4m^2$. The positivity properties \eqref{eqn:dImA} can therefore be continued to finite positive $t$, i.e. away from  the forward limit,
\be
\frac{\pd^n}{\pd t^n}{\rm Im}[A(s,t)] > 0  \;\;\;\; \forall \;\;\;\;  s \ge 4m^2 \, , \, 0 \le t < 4m^2      .
\label{ImAori}
\ee
This is the key property which can be used to derive positivity requirements of the low energy scattering away from the forward scattering limit.

Furthermore, analyticity can be combined with the unitarity condition \eqref{eqn:unitarity_al} to derive the Froissart-Martin bound for the behaviour of the scattering amplitude at fixed $t$ \cite{Froissart:1961ux, Martin:1962rt,Jin:1964zza}
\begin{equation}
\lim_{s \to \infty} |A (s,t)| < C s^{1+\epsilon(t)} ,~~~~0\leq t < 4m^2    ,
\label{eqn:Froissart}
\end{equation}
where $C$ is constant and $\epsilon(t)$ can depend on $t$. The results of \cite{Jin:1964zza} imply that $\epsilon(t)<1$ in the range $0\leq t < 4m^2 $, which in turn implies that the fixed $t$ amplitude may be expressed as a dispersion relation with only two subtractions, just as in the forward limit $t=0$. This is the second key result we shall make use of.
\\

%%%%
\noindent {\bf Dispersion relation:}
We begin with the assumption that the scattering amplitude at fixed momentum transfer, $0 \le t < 4m^2$ (away from the $t=m^2$ pole), is an analytic function of $s$, modulo poles and branch cuts in the usual places. Then by Cauchy's integral formula,
\begin{equation}
A (s,t) = \frac{1}{2 \pi i} \oint_{\cal C} ds'  \; \frac{ A (s',t) }{ s' - s }    ,
\end{equation}
where ${\cal C}$ is a counterclockwise contour inside of which $A$ is analytic. One can deform ${\cal C}$ into an infinite circular contour going around the two branch cuts plus two infinitesimal clockwise circles around the simple poles at $s'=m^2$ and $u(s',t)=m^2$ to obtain
\ba
\label{eq:AmpA1}
A (s,t) =   \frac{\lambda }{m^2 - s} +  \frac{\lambda}{m^2 - u} +  \int_{{\cal C}^{\pm}_\infty} ds'  \frac{  A (s',t) }{ s' - s }   \\
 +  \int_{4m^2}^{\infty}   \frac{d \mu}{\pi}   \( \; \frac{ \text{Im} A (\mu,t) }{ \mu - s } + \frac{ \text{Im} A (\mu,t) }{ \mu - u } \)   \nn  ,
\ea
where ${\cal C}^{\pm}_\infty$ is the semicircle with radius $s' \to\infty$ in the upper/lower half plane, and we have used the Schwarz reflection principle $A (s^*,t) = A^*(s,t)$ to relate the discontinuity along the cuts to the imaginary part of the amplitude, and $s\leftrightarrow u$ crossing symmetry to infer the discontinuity on the left hand cut. Crossing symmetry guarantees that the pole residues $\text{Res}_{u=m^2} A (s, t)=-\text{Res}_{s=m^2} A (s, t)=\lambda$  and for scalar particles these residues are independent of $t$.
The Froissart-Martin bound \eqref{eqn:Froissart} suggests that the contour integrals along ${\cal C}_\infty^\pm$ are not finite in the limit $s' \to\infty$, and so the standard remedy is to perform two subtractions.
In practice, this comes from the identity
\ba
&&   \frac{ \text{Im} A (\mu,t) }{ \mu - s } =  \frac{(s - \mu_p)^2}{(\mu-\mu_p)^2} \frac{ \text{Im} A (\mu,t) }{ \mu - s } \\
&&  +2   \frac{(s - \mu_p)   }{ (\mu-\mu_p)^2 } \text{Im} A (\mu,t) +     \frac{(\mu - s)   }{ (\mu-\mu_p)^2 }\text{Im} A (\mu,t)  \, .\nn
\ea
The subtraction point $\mu_p$
is arbitrary so may be chosen for convenience and can depend on $t$.  By $s \leftrightarrow u$ crossing symmetry, the amplitude \eqref{eq:AmpA1} may be rewritten as
\ba
&& A (s,t) =   a(t) + \frac{\lambda }{m^2 - s} +  \frac{\lambda}{m^2 - u} +  \\
 &&  +\int_{4m^2}^{\infty}   \frac{d \mu}{\pi}   \;   \( \frac{(s - \mu_p)^2 \text{Im} A (\mu,t) }{ (\mu-\mu_p)^2 (\mu - s) } +  \frac{(u - \mu_p)^2 \text{Im} A (\mu,t) }{ (\mu-\mu_p)^2 (\mu - u) } \)\nn ,
\ea
where $a(t)$ absorbs all the remaining integral contributions and is undetermined by analyticity. However by  $t \leftrightarrow s$ crossing symmetry $A(s,t)=A(t,s)$,
we can determine $a(t)$ up to a constant.
Since full crossing symmetry implies that the amplitude must have poles in all three channels, it is convenient to remove these and define
\begin{equation}
B (s,t) = A( s,t) -   \frac{\lambda }{m^2 - s} - \frac{\lambda }{m^2 - u}- \frac{\lambda }{m^2 - t}  \,.
\label{eqn:B}
\end{equation}

For later convenience we choose $\bar \mu_p = -\bar t/2$, where the bar denotes $\bar x := x - 4m^2/3$, redefine the subtraction function $b(t) = a(t) - \frac{\lambda}{m^2 - t}$, and replace $s$ with the variable $v = \bar s +\bar t/2$. In terms of $u$ and $v$, the  $s \leftrightarrow u$ crossing symmetry implies a $v \leftrightarrow  - v$ crossing symmetry. Since $B(s,t)$ is crossing symmetric, it has to be given by an analytic function of $v^2$ in the Mandelstam triangle, $B(s,t) = \tilde B( v (s,t) , t)$ where
\be
 \tilde B ( v , t) = b(t) + \int_{4m^2}^\infty \frac{d \mu}{\pi (\bar \mu + \bar t/2)} \;  \,   \frac{  2 v^2 \, \text{Im} A(\mu,t)  }{ (\bar \mu+\bar t/2)^2- v^2 } \, .
\ee
This is the final form of the dispersion relation that we shall make use of.  \\

%%%%
\noindent{\bf Positivity bounds:}
For $N \ge 1$, we define,
\be
B^{(2N,M)} (t) = \frac{1}{M!} \partial_v^{2N} \partial_t^M  \tilde B( v ,t) \Big|_{v=0} \, ,
\ee
which can be expressed in terms of the positive integrals,
\begin{equation}
I^{(q,p)} (t) = \frac{q!}{p!} \frac{2}{\pi} \int_{4m^2}^\infty \frac{d \mu \;\; \partial_t^p \text{Im} \, A (\mu ,t) }{ (\bar \mu + \bar t /2 )^{q+1}} > 0 ,
\label{eqn:Ipos}
\end{equation}
where we have now made use of the unitarity condition \eqref{ImAori}.
Explicitly,
\begin{equation}
B^{(2N,M)} (t) = \sum_{k=0}^M  \frac{(-1)^k}{k! 2^{k}} I^{(2N+k, M-k)} \,.
\label{eqn:BI}
\end{equation}
The left-hand side is a derivative of the pole subtracted amplitude evaluated at $s \sim m^2$, which can be computed in the LEEFT. This known quantity is related to the integrals $I^{(q,p)}$, which depend on the details of the full UV completion, and are therefore not explicitly known---however, as we have argued, they are required to be positive by unitarity and  analyticity. The goal is now to use \eqref{eqn:BI} to translate the positivity of the integrals \eqref{eqn:Ipos} into a bound on the different derivatives of the low energy amplitudes.

\paragraph{With no $t$ derivatives,} this is straightforward: $B^{(2N,0)} (t) > 0$. This is a generalization of the now familiar constraint on the forward scattering amplitude at $t=0$ \cite{Adams:2006sv}. In particular we note that for $t=0$, then $B^{(2N,0)} (0)$ are just the coefficients in the expansion of the following quantity
\ba
f (s_p) &=& \frac{1}{2 \pi i} \oint_{\cal C'} ds'  \; \frac{ A (s',0) }{ (s' - s_p)^3 }={ \frac{1}{2} \frac{\partial^2 B(s,0)}{\partial s^2}}\Big|_{s=s_p} \\
&=& \sum_{N=1}^{\infty}  \frac{s_p^{2N-2}}{2 (2N-2)!}  B^{(2N,0)} (0) > 0 \, ,\nn
\ea
for $0 \le s_p < 4m^2$, where the contour $\cal C'$ is the same as $\cal C$ but without the circles around the poles. The extension of this bound to $0 \le t < 4m^2$ has previously been noted and used in \cite{Vecchi:2007na, Nicolis:2009qm,Bellazzini:2014waa,Manohar:2008tc,Pennington:1994kc}.

\paragraph{For higher $t$ derivatives,} the $B^{(2N,M)}(t)$ do not immediately satisfy a positivity condition due to the alternating sign structure $(-1)^k$ in \eqref{eqn:BI}. To deal with this, we first note that the various integrals satisfy the inequality,
\ba
I^{(q,p)} < \frac{q}{\mathcal{M}^2}I^{(q-1,p)}\,,
\label{eqn:integral_inequality}
\ea
where $\mathcal{M}^2$ is the minimum value of $\( \bar \mu +\bar t/2 \)$ within the region of integration for $\mu$. For now we could simply set $\mathcal{M}^2=(t+4m^2)/2$, but when dealing with tree-level amplitudes we shall see later that $\mathcal{M}$ may take much larger values.
To see how the previous inequality can be used to our advantage, consider  a single $t$ derivative,
\ba
B^{(2N,1)}=  I^{(2N,1)} - \tfrac{1}{2} I^{(2N+1,0)}
>
 I^{(2N,1)} -\frac{2N+1}{2 \M^2} I^{(2N,0)}\,.\nn
\ea
Since $B^{(2N,0)}=I^{(2N,0)}$, we can immediately infer  that the quantity $Y^{(2N,1)}(t)$ defined as follows
\be
Y^{(2N,1)} = B^{(2N,1)}  + \frac{2N+1}{2 \mathcal{M}^2} B^{(2N,0)}  > I^{(2N,1)} >  0\, ,
\ee
is manifestly positive.

Proceeding onto a second $t$ derivative, we have,
\ba
B^{(2N,2)} = I^{(2N,2)}  - \frac{1}{2} I^{(2N+1,1)} + \frac{1}{8} I^{(2N+2, 0)}  \,.
\ea
Since only one of the terms enters negatively it would be possible to perform just one addition and end up with a quantity which is manifestly positive,
\ba
B^{(2N,2)}+\frac{2N+1}{2\M^2}Y^{(2N,1)}>  I^{(2N,2)}  + \frac{1}{8} I^{(2N+2, 0)}> 0\,.\nn
\ea
 We can however construct a more restrictive bound by performing a second subtraction that also removes the integral   $I^{(2N+2, 0)}=B^{(2(N+1),0)}$ as a result, the following quantity is manifestly positive,
\ba
Y^{(2N,2)}  = B^{(2N,2)}   + \frac{2N+1}{2 {\cal M}^2} Y^{(2N,1)} - \frac{1}{8} B^{(2(N+1), 0)} > 0 \,.\nn
\ea
The previous examples illustrated how to construct positive quantities involving up to two $t$ derivatives of the amplitude. We now present the  general procedure for any number of $t$ derivatives.

Motivated by the previous constructions, we now consider a linear combination of $B$'s. On dimensional grounds, if we want to be dealing with up to $M$ $t$ derivatives and $2N$ $v$ derivatives, then one should consider
\ba
\label{eq:sumB}
\sum_{r=0}^{M/2} c_r B^{(2N+2r,M-2r)} = \sum_{k=0}^{M/2}\alpha_k I^{(2N+2k,M-2k)}   \\
 -  \sum_{k=0}^{(M-1)/2} (-1)^k \beta_k I^{(2N+2k+1,M-2k-1)} \,,\nn
\ea
where for each $B$, we have split its sum \eqref{eqn:BI} into odd and even parts, and we have defined
\ba
\label{eqn:cbark}
\alpha_k=\sum_{r=0}^k \frac{2^{2(r-k)} c_r }{ (2k-2r)! } \, , \,
\beta_k= (-1)^k \sum_{r=0}^k \frac{2^{2(r-k)-1} c_r }{ (2k-2r+1)! }\,.\qquad
\ea
 Now the  coefficients $c_r$ can be chosen precisely so as to remove every integral $I^{(2q,p)}$  with $p < M$ on the first line of \eqref{eq:sumB}, i.e. so that $\alpha_k=\delta_{k,0}$.  This requirement sets % the coefficients to be
\be
c_0 = 1 \quad{\rm and}\quad c_k = - \sum_{r=0}^{k-1} \frac{ 2^{2 (r-k)} c_r}{ (2k-2r)! }\,, \quad \forall \quad k\ge 1\,.
\label{eqn:ck}
\ee
These coefficients are easily computed to any  desired order, and one can check that $\beta_k \ge 0$. It follows that the following quantity is  manifestly positive,
\ba
&&\sum_{r=0}^{M/2} c_r B^{(2N+2r,M-2r)} + \sum_{k \; \text{even}}^{(M-1)/2}  \beta_k   I^{(2N+2k+1,M-2k-1)}  \nn  \\
&&= I^{(2N,M)} +  \sum_{k \; \text{odd}}^{(M-1)/2}  \beta_k    I^{(2N+2k+1,M-2k-1)} > 0  .
\ea
Now using the integral inequalities \eqref{eqn:integral_inequality}, we can recursively construct the $Y^{(2N,M)}$'s as follows
\ba
\begin{split}
\label{eqn:Y}
& Y^{(2N,M)} = \sum_{r=0}^{M/2} c_r B^{(2N+2r,M-2r)}  \\
&+ \frac{1}{{\cal M}^2} \sum_{k \; \text{even}}^{(M-1)/2} (2(N+k)+1)  \beta_k   Y^{(2(N+k),M-2k-1)}> 0\,,
\end{split}
\ea
where $c_r$ and $\beta_k$ are given by \eqref{eqn:cbark} and \eqref{eqn:ck} and for now $\mathcal{M}^2=(t+4m^2)/2$.
By construction, $Y^{(2N,M)}(t)\ge I^{(2N,M)}> 0$ for all $N\ge 1$, all $M\ge 0$ and for any $0 \le t < 4m^2$.

As a result, we have successfully constructed combinations of derivatives of the LEEFT amplitude, $Y^{(2N,M)}$, which must be positive if there were to ever exist a local, analytic and Lorentz invariant UV completion. Although we have used $N \geq 1$ in the derivation, note that $s \leftrightarrow t$ crossing means that $Y^{(q,p)}$ and $Y^{(p,q)}$ are not truly independent quantities, and so our list of bounds is exhaustive (as we have included the case $M=0$, if not $N=0$). \\

%%%%
\noindent{\bf Tree level bounds:}
The previous positivity bounds were placed on the exact {\it all loop} scattering amplitude. It is for this reason that the threshold on the $\mu$ integrals was taken to be $\mu=4m^2$, associated with elastic scattering, e.g. one-loop processes with two intermediate scalars. However, a more practical use of these bounds is to bound coefficients in the tree-level Lagrangian of the LEEFT. In doing this, there is a fundamental difference in approach since at tree level in the low energy effective theory, nonzero ${\rm Im}A(s,t)$  can only arise at and above the threshold $\mu =\Lambda_{\rm th}^2$ defined as the mass of the first state that lies outside of the effective theory, i.e. the cutoff of the LEEFT. Contributions in the region $4m^2 \le \mu < \Lambda_{\rm th}^2$ come from loops of light particles already included in the LEEFT and so will not show up in the tree level amplitude. At a pragmatic level this means that since $\Lambda_{\rm th}^2 \gg 4m^2$, the lower limit of the $\mu$ integrals may be taken to be $\Lambda_{\rm th}^2$ and we may make use of stronger integral inequalities \eqref{eqn:integral_inequality} with $\mathcal{M}^2 = \Lambda_{\rm th}^2$,
and define $Y^{(2N,M)}_{\rm tree}$ through \eqref{eqn:Y} with $\mathcal{M}^2 = \Lambda_{\rm th}^2$, where $B$ is now understood to be the pole subtracted amplitude computed \emph{to tree level only} in the LEEFT.
The tree level bounds are then
\be
 Y^{(2N,M)}_{\rm tree} (t, \Lambda_{\rm th}) > 0 \, ,
\ee
and may in some cases provide additional constraints on $\Lambda_{\rm th}$ in terms of the LEEFT coefficients and the mass $m$.  \\

%%%%
\noindent{\bf Matching against the low energy effective theory:}
To see the power of these bounds, consider a general effective theory tree amplitude, with poles subtracted $B(s,t)$. Any such amplitude, may be given by an analytic function of the crossing symmetric variables
\be
x=-(\bar s \bar t + \bar t \bar u+ \bar u \bar s ) \quad {\rm and }\quad y =- \bar s \bar t \bar u \, ,
\ee
with $\bar s+\bar t + \bar u =0$, so that
\be
B(s,t) =  \sum_{nm} \frac{a_{nm}}{\Lambda^{4n+6m}}x^n y^m \, ,
\ee
where $\Lambda$ is some theory specific scale introduced to make the $a_{nm}$ dimensionless. Note that $x= v^2 + \frac{3}{4} \bar t^2$ and $y =\bar t  v^2 - \frac{1}{4} \bar t^3$.
For concreteness, in what follows we evaluate the derivatives of $B$ at $t=4m^2/3$, corresponding to the maximally crossing symmetric point $s=t=u$.
Then the derivatives $B^{(2N,M)}$ in terms of the $a_{nm}$ are
\be
B^{(2N,M)}=(2N)! \sum_q \frac{d_q}{\Lambda^{4N+2M}} a_{N-M+3q,M-2q} \, ,
\ee
where $q! d_q= (3/4)^q (3q+N-M) {}_2F_1(-q,2q-M,1-M+N+2q,-1/3)$.
Up to eighth order in energy, we  have  four EFT coefficients $(a_{00}, \; a_{10}, \; a_{01}, \; a_{20})$,
which are bounded by,
\ba
Y^{(2,0)} : \;\;\;\; a_{10} &>& 0 \\
Y^{(2,1)} : \;\;\;\; a_{01} &>& - \frac{3 \Lambda^2}{2 \Lambda_{\rm th}^2 } a_{10} \\
Y^{(4,0)} : \;\;\;\; a_{20} &>& 0\,,
\ea
and the bound provided by $Y^{(2,2)}$ ends up being automatically satisfied by the previous ones.
These bounds clearly distinguish two separate cases. If for a particular theory $a_{01}$ were found to be positive, then the bounds are satisfied provided $a_{10},a_{20}>0$ regardless of $\Lambda_{\rm th}$. However, if for a theory $a_{01}$ is negative, then for that theory we can put an upper bound on the threshold $\Lambda_{\rm th}$,
\be
 \Lambda_{\rm th}^2 \le \frac{3a_{10}}{2 |a_{01}|} \Lambda^2 \, .
\ee
This bound on the cutoff of the theory is logically separate from the scale at which perturbative unitarity is broken. If we build an EFT from the bottom up, without initial knowledge of the true scale $\Lambda_{\rm th}$, then we can separately compute $\Lambda_{\rm pert}$, the scale at which perturbative unitarity is violated, and if $a_{01}<0$, $\Lambda_{\rm th}$ the scale at which analyticity is broken. If $\Lambda_{\rm th}$ is found to be lower than $\Lambda_{\rm pert}$ then this implies we should have added irrelevant operators suppressed by $\Lambda_{\rm th}$.

\vskip 5pt

%%%%
\noindent{\bf Discussion:}
 In this work, we have paved the way towards constraining all LEEFTs in a novel way and have found an infinite number of new bounds that remain valid even away from the forward scattering limit. Violating {\it any} of this infinite number of bounds directly implies the absence of any possible local, unitary and Lorentz invariant UV completion for the scalar field theory. These place bounds on the independent coefficients in the EFT Lagrangian. As a by-product we have shown how, together with the assumption of analyticity, the scale of new physics can be further constrained, beyond what would be possible from using perturbative unitarity alone. In certain cases, it is conceivable that the stricter requirement of analyticity implies a lower cutoff on the EFT than would be implied by standard perturbative unitarity considerations, a simple example of which will be given in \cite{dRMTZ2:2017}. These results are derived under the assumption of a mass gap, principally to make use of the Froissart-Martin bound, and to ensure the existence of the analytic Mandelstam triangle.  An extension of these positivity bounds to particles with nonzero spin will be discussed in \cite{dRMTZ:2017}.

\noindent {\bf Acknowledgments:}
CdR thanks the Royal Society for support at ICL through a Wolfson Research Merit Award. SM is funded by the Imperial College President's Fellowship. AJT thanks the Royal Society for support at ICL through a Wolfson Research Merit Award

\bibliography{references}

\end{document}